\documentclass{Rinton-P9x6}
\usepackage{graphics}

\begin{document}

\title{New Challenges and Directions in Casimir Force Experiments}

\author{Davide Iannuzzi, Ian Gelfand, Mariangela Lisanti, and Federico Capasso}

\address{Division of Engineering and Applied Sciences, Harvard University, Cambridge, MA 02138, USA
\\E-mail: capasso@deas.harvard.edu
}

\maketitle

\abstracts{This article is divided in three sections. In the first section we briefly review some high precision experiments on the Casimir force, underlying an important aspect of the analysis of the data. In the second section we discuss our recent results in the measurement of the Casimir force using non-trivial materials. In the third section we present some original ideas for experiments on new phenomena related to the Casimir effects.}

\section{Past measurements of the Casimir force: discussion of the experiments with a 1\% claimed accuracy}

Two years ago an experiment at Bell Labs showed that MicroElectroMechanical Systems (MEMS) can be actuated by means of the Casimir force~\cite{chan1}. The effect was demonstrated by bringing a polystyrene sphere close to a micromachined torsional device that consists of a polysilicon plate suspended over the substrate by means of two thin rods. Both the surfaces were coated with a layer of gold and kept at the same potential to cancel out the electrostatic interactions. The rotation angle of the microtorsional device induced by the Casimir attraction was measured as a function of the distance between the sphere and the plate. The agreement of the experimental result with the theoretical calculations was in the 1\% range.  

In spite of the apparent excellent accuracy of the experiment, it was underlined that the high level of agreement with the theory could be fortuitous. In fact the calculations are strictly correct only for bulk materials, while in the experiment both the surfaces were obtained by evaporating thin metallic layers over non-metallic substrates.\footnote{It is widely known that the dielectric function of evaporated films can differ from those of the bulk material.} Furthermore, the Casimir force strongly depends on the roughness of the two surfaces. Therefore, small errors in the evaluation of the roughness could give rise to large corrections to the theoretical result. Both the effects are potential causes of errors larger than 1\%.~\cite{lam1,lam2}

\begin{figure}[t]
\resizebox{\textwidth}{!}{\rotatebox{-90}{\includegraphics{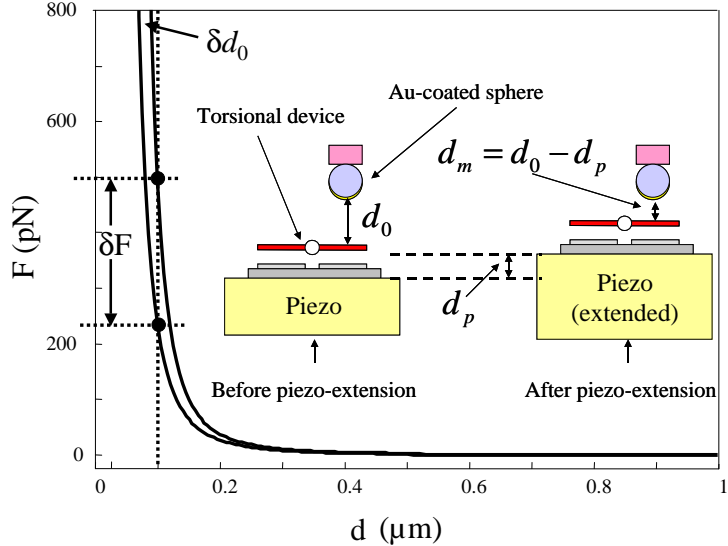}}}
\vspace{-35pt}
\caption{Effect of the error on the distance on the measurement of the Casimir force. Due to the strong dependence of $F$ on $d_m$, small errors in $d_0$ give rise to large errors in $F$.\label{fig:ddd}}
\end{figure}

However, there is another important contribution to the systematic error that was not explicitly discussed in reference $1$ that is extremely important for all the high precision measurements of the Casimir effect. The Casimir force between a sphere and a plate, $F$, strongly depends on the distance of the surfaces $d_m$. For ideal metals, one obtains $F \propto d_m^{-3}$. For real materials, the dependence of $F$ on $d_m$ is more complicated, but still very strong. In the Bell Labs experiment, the micromachined device was mounted on a calibrated piezoelectric stage that could be extended to reduce the distance between the plate and the sphere (see figure \ref{fig:ddd}). $d_m$ could be obtained as $d_0-d_p$, where $d_p$ is the extension of the piezoelectric stage, which was known with subnanometric precision, and $d_0$ is the distance between the sphere and the plate when the piezoelectric stage was not extended.\footnote{In the context of this discussion, we can neglect the effect of the rotation of the plate on $d_m$. For a detailed analysis of this point, see reference 1} However, $d_0$ was not known {\it a priori}. Remarkably, small errors in $d_0$ can give rise to large errors on the measurement, as shown in figure \ref{fig:ddd}. To determine $d_0$ in the Bell Labs experiment the Casimir force data were fitted using the theoretical curve, keeping $d_0$ as a free parameter. The error on $d_0$ arising from the fit was equal to $1$ nm. For $d_m$ smaller than $\simeq 300 $ nm, the force at a distance $d_m$ differs for more than 1\% from the force at a distance $d_m+1$ nm . Therefore, it does not make any sense to compare the data and the theory at the less than 1\% level at short distances. 

The $d_0$ problem was actually firstly noticed by T. Ederth,~\cite{ederth} who measured the Casimir force between two large, atomically flat, crossed cylinders. Also in that case, one of the two cylinders was mounted on a piezoelectric stage, and the distance between the two surfaces was determined by $d_m=d_0-d_p$. Since the surfaces were atomically flat, $d_0$ could be determined by measuring $d_p$ when the two surfaces came to contact ($d_m=0$). However, the deformation of the cylinders for $d_m=0$ turned out not to be negligible and, actually, caused large systematic errors in the measurement. 

Interestingly, the $d_0$ problem was not explicitly considered in the recent experiments on the Casimir force with Atomic Force Microscopy (AFM).~\cite{moh1} In that apparatus, a gold-coated polystyrene sphere, mounted on the cantilever of an AFM, was brought close to a metallic surface, and the deflection of the cantilever induced by the Casimir attraction was measured as a function of the distance between the two surfaces. The authors claimed an experimental uncertainty of 3.5 pN at the shortest distance $d_m=62$ nm, corresponding to an agreement with the theory better than 1\%. However, it must be noted that the force at $d_m=62$ nm differs from the force at $d_m=62+\delta$ by more than 3.5 pN (the experimental uncertainty claimed by the authors) when $\delta$ is larger than a few angstroms. This implies that $d_0$ should be measured with atomic precision because otherwise the errors on $d_m$ would be more relevant than the experimental uncertainty in the measurement of $F$. However, as the authors reported, the error on $d_0$ was of the order of $\pm 1$ nm, and thus should not be neglected in the analysis of the experimental errors. For this reason the precision of the experiment was overestimated.

In order to illustrate how important the $d_0$ problem is, we analyzed the data recently reported by R. Decca and his collaborators.~\cite{decca} With an experimental apparatus similar to the one used at Bell Labs, they measured the Casimir force between a gold and a copper film. Interestingly, they observed discrepancy of the data from the theoretical predictions of the order of $1\%$. The disagreement was attributed to the lack of a complete characterization of the physical properties of the materials that are needed to calculate exactly the Casimir force - namely, the dielectric function and the surface roughness. However, an analysis similar to the one above shows that the discrepancy can be almost completely removed if all the data are shifted of 1 nm, consistently with an error of 1 nm on $d_0$.

In light of what we have discussed so far, we want to emphasize that experiments on the Casimir effect with 1\% precision can be achieved only if the distance between the attracting surfaces is known with a very high level of accuracy. In the experiments measuring the attraction between a sphere and a plate at distances of $\simeq 100$ nm, errors of a few angstroms are already too large to allow a meaningful comparison of the data with the theory at the 1\% level. 

\section{Measurements of the Casimir force using non-trivial materials}

Our group at Harvard University is interested in experiments exploring new aspects of the Casimir effect. Recently, for instance, we measured the Casimir force between a gold surface and a Hydrogen Switchable Mirror (HSM).

\begin{figure}[t]
\resizebox{\textwidth}{!}{\rotatebox{-90}{\includegraphics{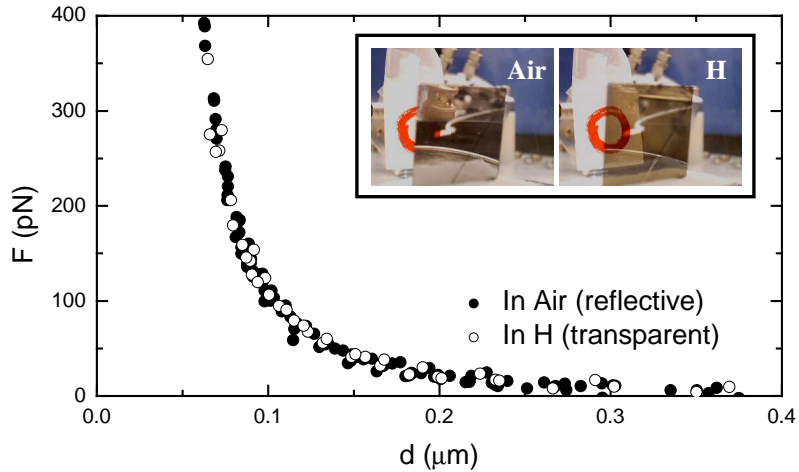}}}
\vspace{-75pt}
\caption{Measured Casimir force between a gold plate and a sphere coated with a HSM in air and in a Hydrogen rich atmosphere. In the inset: a HSM in air (left) and the same mirror in a Hydrogen rich atmosphere (right). \label{fig:hsm}}
\end{figure}

HSMs~\cite{griessen} are shiny metals in their {\it as deposited} state. However, when they are exposed to a Hydrogen rich atmosphere, they become optically transparent (see inset of figure \ref{fig:hsm}). Since the Casimir force between two surfaces depends on their optical properties, the Casimir attraction between two HSMs in their {\it as deposited} state should be different from the Casimir attraction between the same HSMs immersed in a Hydrogen atmosphere.

To verify this prediction, we used the Bell Labs experimental apparatus,~\cite{chan1} this time coating the sphere with a HSM (Ni-Mg-Pd).~\cite{rich} We measured the Casimir force between the sphere and the torsional device in air (reflective state) and in a Hydrogen rich atmosphere (transparent state). The result is reported in figure \ref{fig:hsm}. No change of the force was observed in the two different states. 

This counterintuitive result can be explained using the Lifshitz equation.~\cite{lifshitz} The dielectric function of the HSM are known only in a limited range of wavelengths (from $\simeq 0.3$ $\mu$m to $\simeq 2.5$ $\mu$m). In this interval, upon hydrogenation,  the reflectivity changes from $\simeq 80\%$ to $\simeq 10\%$, while the transparency increases from almost 0 to $\simeq 20\%$. Outside this interval, there are no available data. In order to have significant changes in the Casimir force at $\simeq 100$ nm distance, the dielectric properties must change over a large wavelength range (from few tens of nanometers up to few hundreds of $\mu$m). This is in contrast with the common belief that only the modes with a wavelength comparable to the distance between the surfaces contribute to the Casimir force, which can be easily verified with the Lifshitz equation~\cite{bo}. Therefore, if the dielectric properties of the HSMs do not change outside the $0.3$ $\mu$m-$2.5$ $\mu$m range, the change in the Casimir force upon hydrogenation could be too small to be measurable.

However, it is interesting to note that HSMs are always covered with a thin layer of Pd ($\simeq 50$ \AA). The contribution of this layer to the Casimir force is not known and cannot be calculated. We made some preliminary measurements of the Casimir attraction between thin films and observed that thin films can give rise to the largest contribution to the force. Therefore, in the case of the experiment with the HSMs, the thin layer of Pd deposited on top of the film might give rise to a dominating large force. Since the properties of the Pd layer do not change in Hydrogen, the force observed in air and the force observed in a Hydrogen rich atmosphere are the same. Further investigations are under way to confirm this result.
 
\section{Some ideas for the investigation of new phenomena related to the Casimir effect}

\subsection{Casimir torque}

\begin{figure}[t]
\resizebox{\textwidth}{!}{\rotatebox{-90}{\includegraphics{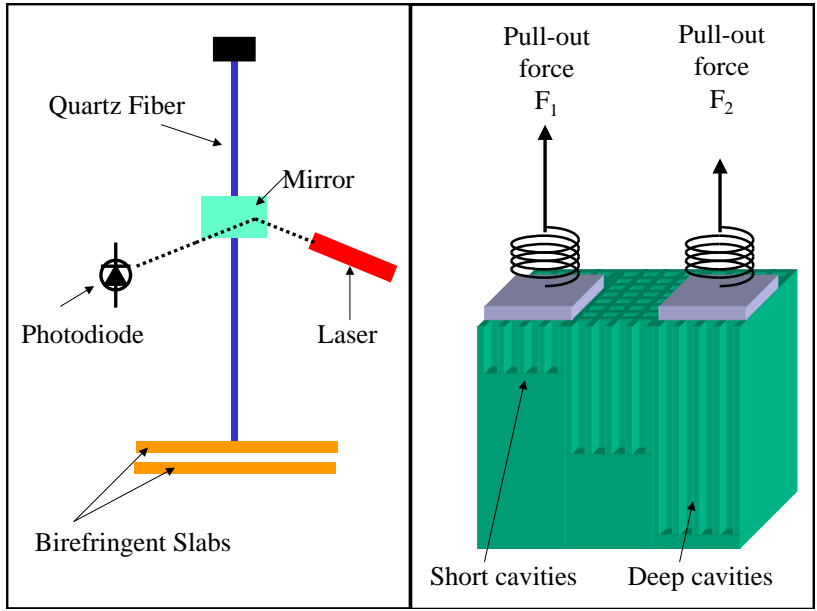}}}
\vspace{-20pt}
\caption{Left: a possible set up for the measurement of the Casimir torque between two birefringent slabs. Right: a schematic view of a set up for the measurement of the repulsive Casimir force in rectangular cavities.\label{fig:new}}
\end{figure}

In figure \ref{fig:new}, we present a possible scheme for the measurement of the Casimir torque between two parallel birefringent plates. The torque arises from the fact that the vacuum energy of the system depends on the angle between the principal axis of the two slabs. We have calculated the Casimir torque using the theory developed by Y. Barash~\cite{barash} for the case of two LiNbO$_3$ surfaces.\footnote{We have used the simplified equation (34) of reference $11$, which is valid only in the non-retarded limit and for materials with relatively small birefringence characteristics. However, large deviations from this rough estimation are not expected for a more complete calculation.} For two disks of radius 1 cm, kept at a distance of 1 $\mu$m, the Casimir torque is equal to $\sim 10^{-15}$ Nm. The experimental apparatus shown in figure \ref{fig:new} has a sensitivity of $\sim 10^{-17}$ Nm,~\cite{beth} which should allow us to measure the effect. The main problems of the measurement are dust and parallelism between the surfaces. We believe, however, that dust can be removed using CO$_2$ snow cleaning and parallelism can be controlled with an interferometric system.  

\subsection{Topological repulsive Casimir force}

In 1968, T.H. Boyer~\cite{boyer} showed that for a perfectly conducting spherical shell the Casimir effect should give rise to an outward pressure. Similar {\it repulsive} Casimir forces are expected to also take place in cubic cavities and some rectangular cavities.~\cite{maclay} These theoretical results have never been experimentally investigated.

In figure \ref{fig:new}, we show a possible set up to measure the repulsive Casimir force in rectangular cavities. The idea is to measure the pull-out force necessary to remove a metallic plate from an array of cavities as a function of the cavity depth. Comparing the data obtained from different cavities, one can extract, by difference, the repulsive contribution arising from the Casimir effect. 

\section*{Acknowledgments}
This project was partially supported by the National Science Foundation under the Grant No. PHY-0117795.


\begin{thebibliography}{99}

\bibitem{chan1} H.B. Chan, V.A. Aksyuk, R.N. Kleiman, D.J. Bishop, and F. Capasso, {\it Science} {\bf 291} 1941 (2001)

\bibitem{lam1} S.K. Lamoreaux, \Journal{\PRL}{83}{3340}{1999}

\bibitem{lam2} S.K. Lamoreaux, {\it Phys. Rev.} A {\bf 59}, R3149 (1999)

\bibitem{ederth} T. Ederth, {\it Phys. Rev.} A {\bf 62}, 062104 (2000)

\bibitem{moh1} B.W. Harris, F. Chen, and U. Mohideen, {\it Phys. Rev.} A {\bf 62}, 052109 (2000)

\bibitem{decca} R.S. Decca, D. Lopez, E. Fischbach, and D. E. Krause, \Journal{\PRL}{91}{050402}{2003}

\bibitem{griessen} J.N. Huiberts {\it et al.}, {\it Nature} {\bf 380} {231} {1996}

\bibitem{rich} T.J. Richardson {\it et al.}, {\it Appl. Phys. Lett.} {\bf 78}, 3047 (2001)

\bibitem{lifshitz} E.M. Lifshitz, {\it Sov. Phys. JETP} {\bf 2}, 73 (1948)

\bibitem{bo} M. Bostr\"{o}m and B.E. Sernelius, {\it Phys. Rev.} A {\bf 61}, 046101 (2000)

\bibitem{barash} Y. Barash, {\it Izvestiya vuzov, Radiofizika} {\bf XXI-11}, 1637 (1978)

\bibitem{beth} R.A. Beth, {\it Phys. Rev.} {\bf 50}, 115 (1936)

\bibitem{boyer} T.H. Boyer, {\it Phys. Rev.} {\bf 174}, 1764 (1968)

\bibitem{maclay} G.J. Maclay, {\it Phys. Rev.} {\bf A61}, 052110 (2000)



\end{thebibliography}
\end{document}